\documentclass[twocolumn,showpacs,pre,amsmath,amssymb]{revtex4-2}
\usepackage{xcolor,epsfig,bm,amsmath,amssymb}
\newcommand{\cN}{{\cal N}} 
\newcommand{\cO}{{\cal O}}

\newcommand{\cS}{{\cal S}}

\newcommand{\E}      {{\mathbb{E}}}                 
\newcommand{\R}      {{\mathbb{R}}}                 
\newcommand{\Z}      {{\mathbb{Z}}}                 
\newcommand{\bbbone}    {{\mathbb{I}}}                  

\newcommand{\rd} {\mathrm d}

\newcommand{\re} {\mathrm e}

\newcommand{\va} {{\bm a}}
\newcommand{\vb} {{\bm b}}
\newcommand{\vc} {{\bm c}}

\newcommand{\vu} {{\bm u}}
\newcommand{\vv} {{\bm v}}
\newcommand{\vw} {{\bm w}}
\newcommand{\vx} {{\bm x}}

\newcommand{\nn}{\nonumber} 
\newcommand{\be}{\begin{equation}} 
\newcommand{\ee}{\end{equation}}  
\newcommand{\bea}{\begin{eqnarray}}
\newcommand{\eea}{\end{eqnarray}}
\newcommand{\barr}{\begin{array}}
\newcommand{\earr}{\end{array}}
\newcommand{\bcent}{\begin{center}} 
\newcommand{\ecent}{\end{center}}  
\newcommand{\bit}{\begin{itemize}}
\newcommand{\eit}{\end{itemize}}
\newcommand{\bq}{\begin{quote}}
\newcommand{\eq}{\end{quote}}

\begin{document}
\title{Gaussian Level-Set Percolation on Complex Networks}
\author{Reimer K\"uhn}
\affiliation{Department of Mathematics, King's College London, Strand, London WC2R 2LS,UK}
\date{\today}


\begin{abstract}
We present a solution of the problem of level-set percolation for multivariate Gaussians defined in terms of weighted graph Laplacians on complex networks. It is achieved using a cavity or message passing approach, which allows one to obtain the essential ingredient required for the solution, viz. a self-consistent determination of locally varying percolation probabilities. The cavity solution can be evaluated both for single large instances of locally tree-like graphs, and in the thermodynamic limit of random graphs of finite mean degree in the configuration model class. The critical level $h_c$ of the percolation transition is obtained through the condition that the largest eigenvalue of a weighted version $B$ of a non-backtracking matrix satisfies $\lambda_{\rm max}(B)|_{h_c} =1$. We present level-dependent distributions of local percolation probabilities for Erd\H{o}s-R\'enyi networks and and for networks with degree distributions described by power laws. We find that there is a strong correlation between marginal single-node variances of a massless multivariate Gaussian and local percolation probabilities at a given level $h$, which is nearly perfect at negative values $h$, but weakens as $h\nearrow 0$ for the system with power law degree distribution, and generally also for negative values of $h$, if the multivariate Gaussian acquires a non-zero mass. The theoretical analysis simplifies in the case of random regular graphs with uniform edge-weights of the weighted graph Laplacian of the system and uniform mass parameter of the Gaussian field. An asymptotic analysis reveals that for edge-weights $K=K(c)\equiv 1$ the critical percolation threshold $h_c$ decreases to 0, as the degree $c$ of the random regular graph diverges. For $K=K(c)=1/c$, on the other hand, the critical percolation threshold $h_c$ is shown to diverge as $c\to\infty$.
\end{abstract}

\maketitle
\section{Introduction}
The present paper is about the problem of Gaussian level-set percolation on complex networks. It is concerned with the distribution of sizes of contiguous clusters over which a multivariate Gaussian defined on the vertices of a network exceeds a given level $h$, and in particular with the question whether, for a given level $h$, there exists a giant connected component of such vertices (referred to as $h$GCC in what follows) that occupies a finite fraction of vertices in the large system limit. Intuitively, one would expect that there would always exist a {\em finite\/} critical level $h_c$, above which all contiguous clusters for which the Gaussian exceeds the level $h \ge h_c$ would be bounded, whereas for $h < h_c$ there would be an $h$GCC on which the components of the Gaussian exceed the level $h$. Key objectives of the Gaussian level-set percolation problem then are (i) to find the critical level $h_c$ for a given network and a given multivariate Gaussian defined on it, and (ii) to determine the fraction of nodes in the $h$GCC for $h<h_c$. A moment of reflection shows that for complex networks with non-degenerate degree distributions, and thus usually locally varying Gaussian marginal densities, the probabilities of vertices of a network to belong to an $h$GCC, would {\em themselves\/} be locally varying (for $h < h_c$). It turns out that analytic control over the locally varying local percolation probabilities is key to a full solution of the problem. 

Level-set percolation of multivariate distributions is to be contrasted with the simpler case of independent Bernoulli percolation which is concerned with the question whether --- as a function of the relative density of existing links in a lattice or a graph, either by construction or after random {\em independent\/} removal of a subset of edges or vertices --- the system either decomposes into a collection of finite clusters, or on the contrary exhibits a connected component (GCC) that occupies a finite fraction of vertices in the large system limit (see e.g.,\cite{Ess80,Kes82}). Because of the probabilistic independence of vertex or edge removal in the case of standard independent Bernoulli percolation, the problem is much simpler than the case of level-set percolation of random fields, as correlations between values of a multivariate distribution at different vertices of a graph imply that level-set percolation is in fact a version of {\em correlated\/} Bernoulli vertex percolation. The correlations that are present in the case of level-set percolation are the key ingredient that make the problem so much harder than independent Bernoulli percolation. They also entail that the phenomenology of the problem is richer and may well defy simple intuition built on independent Bernoulli percolation. Indeed, early on Molchanov and Stepanov \cite{MoSt83a,MoSt83b} disproved, for instance, the naive intuition mentioned above, according to which there  would {\em always\/} exist a finite critical level $h_c$ separating the percolating phase from the non-percolating phase.

Level set percolation has in the past been studied in the continuum ($\R^d$) \cite{MoSt83a, MoSt86, MuiVan20,CaoSan21, MuiSev22, Mui22, MuRiVa23}, on hypercubic lattices ($\Z^d$) \cite{MoSt83b, BLM87, RoSz13, DrPrRo18, Mui22}, and on random graphs \cite{AbSz18, AbCe20a, AbCe20b, DrPrRo22, CoKe23, DrPrRo23, Cer23, CeLo23}. Because of the correlations in the problem much less is known about it than about standard Bernoulli percolation. The additional difficulty created by these correlations notwithstanding, a number of key properties of Gaussian level-set percolation have been established over the years. Among them are existence (e.g. \cite{MoSt83a,MoSt83b,MoSt86,BLM87,RoSz13,AbSz18,Mui22,MuRiVa23}) and sharpness \cite{Mui22} of the percolation transition. In the case of random graphs, uniqueness \cite{AbCe20b} and extensivity of the $h$GCC \cite{CoKe23,Cer23,CeLo23} in the percolating phase, as well as critical exponents of the transition on transient graphs \cite{CeLo23,DrPrRo23} have recently been added to the set of known results. However, a lot still remains open. In particular, explicit solutions of the problem that go beyond a characterization of the near critical regime \cite{CeLo23, DrPrRo23} have to the best our knowledge until recently been unavailable, in marked contrast to the case of independent Bernoulli percolation on  random graphs, for which full solutions are meanwhile textbook material (e.g., \cite{AB02, DM03, NewBk10}). In very recent shorter version of the present paper \cite{Ku24a}, we managed to fill exactly this gap. The purpose of the present paper is to expand on the details of the solution, and to present and discuss a number of additional results.

The remainder of the paper is organized as follows. In Sect.\,\ref{secDefGLS} we introduce Gaussian level-set percolation in formal terms and describe its analysis for single large instances of locally tree-like graphs using a cavity or message-passing approach. This includes obtaining the critical percolation threshold $h_c$ for a given problem through a stability analysis of the cavity equations as well as an asymptotic analysis of locally varying percolation probabilities in the vicinity of the percolation transition. Section \ref{secThDLim} describes the probabilistic self-consistency argument that allows one to obtain a solution of the level-set percolation problem in the thermodynamic limit of infinite system size for random graphs in the configuration model class. Simplified equations for macroscopic level dependent percolation probabilities are obtained for the special case of random regular graphs. We present and analyze our main results in Sect.\,\ref{secRes}. Section \ref{secSummary} finally contains a summary and discussion. Some technical parts of the theory are relegated to appendices.

\section{Gaussian Level-Set Percolation on Complex Networks}
\label{secDefGLS}
\subsection{The Problem}
We consider a (random) graph of $N$ vertices $i = 1, 2,\dots, N$, on which a multivariate Gaussian is defined via
\be 
P(\vx) =\frac{1}{Z}\,\re^{-\cS(\vx)}\ ,
\label{GFF}
\ee 
in which $\cS(\vx)$ quadratic in the $x_i$, i.e.,
\be
\cS(\vx) = \frac{1}{2}\sum_i \mu_i x_i^2+\frac{1}{4}\sum_{i,j} K_{ij}(x_i-x_j)^2
\label{H}
\ee
with $\mu_i \ge 0$ and $K_{ij}=K_{ji} >0$ for vertices of the network connected by an edge, and $K_{ij}=K_{ji} \equiv 0$ otherwise. The $\mu_i$ are referred to as mass parameters, which, unlike in the shorter version \cite{Ku24a} of this paper, we allow to be locally varying.  For $\mu_i \equiv 0$ the field is referred to as massless. The positivity of the non-zero edge weights is needed to keep the Gaussian normalizable also in the massless case $\mu_i\equiv 0$. In Eq.\,\eqref{H}, the first sum is over all $N$ vertices of the graph, while both sums in the double sum are over all $N$ vertices of the graph. The double sum in Eq.\,\eqref{H} is a quadratic form of a weighted graph Laplacian $\Delta^{(K)}$ with edge weights $\{K_{ij}\}$, i.e.
\be
\frac{1}{4}\sum_{i,j} K_{ij}(x_i-x_j)^2 = -\frac{1}{2} \sum_{i,j} \Delta^{(K)}_{ij}\,x_i x_j\ .
\ee
The analysis of level-set percolation on random graphs {\em requires\/} evaluating locally varying {\em node dependent probabilities\/} of vertices of the graph to belong to the $h$GCC. This is achieved by adapting an approach developed in \cite{KNZ14,KuRog17}. It is based on cavity or message passing ideas specifically designed to analyse problems on locally tree-like graphs. As for the time being we are only interested in evaluating heterogeneous percolation probabilities rather than also cluster size distributions, a simpler version described, e.g., in \cite{ShirKaba10, BTB+21} can be used. 

\subsection{Single Instance Cavity Analysis}
For a node $i$ to belong to the $h$GCC, the multivariate Gaussian at $i$ must itself exceed the specified level, i.e., $x_i\ge h$, {\em and\/} it must be connected to the $h$GCC via at least one of its neighbors. Introducing indicator variables $n_i\in \{0,1\}$ which signify whether $i$ is ($n_i=1$) or is not ($n_i=0$) in the  $h$GCC, one can express this logic as
\be 
n_i = \chi_{\{x_i\ge h\}}\,\Big(1 - \prod_{j\in \partial i} (1 - \chi_{\{x_j\ge h\}} n_j^{(i)}\big)\Big)\ .
\label{ni}
\ee 
In this equation the first factor, i.e., the characteristic function $\chi_{\{x_i\ge h\}}$, expresses the fact that the component $x_i$ of the multivariate Gaussian must itself exceed the specified level, while the second factor expresses the fact that $i$ is connected to the hGGC via at least one neighbor. This in turn requires that for at least one $j\in \partial_i$ the multivariatee Gaussian must exceed the specified level ($x_j \ge h$), {\em and\/} that it must be connected to the $h$GCC via one of {\em its\/} neighbors other than $i$ (on the cavity graph $G^{(i)}$ from which $i$ and the edges connected to it are removed); this is expressed by the cavity indicator variable $n_j^{(i)}$  taking the value $n_j^{(i)}=1$.

For the cavity indicator variable $n_j^{(i)}$ to take the value 1, it is required that on $G^{(i)}$ the node $j$ is itself connected to the $h$GCC via at least one of its neighbors (other than $i$). This is expressed by the condition
\be
n_j^{(i)} = 1 - \prod_{\ell\in \partial j\setminus i} \big(1- \chi_{\{x_\ell\ge h\}} n_\ell^{(j)}\big)\ .
\label{nji}
\ee
Equations \eqref{nji} constitute a set of self-consistency equations (one equation for each edge $(i,j)$ of the graph) that allow one to self-consistently determine the values of the $\{n_j^{(i)}\}$ for a given graph and a given realization of the multivariate Gaussian. Solutions are very efficiently obtained by forward iteration from random initial conditions. From the solution, one can then evaluate the $n_i$ from Eq.\,\eqref{ni}. They indicate whether nodes are or are not on the $h$GCC of the network for that realization of the multivariate Gaussian 

Node dependent percolation {\em probabilities\/} at level $h$ are obtained by averaging Eqs.\,\eqref{ni} over all realizations of the multivariate Gaussian $\vx$ with joint PDF described by Eqs.\,\eqref{GFF} and \eqref{H}, i.e., by evaluating $g_i = \E_\vx[n_i]$ for all vertices $i$ of the graph, in which $\E_\vx[\cdot]$ denotes an expectation evaluated over the multivariate Gaussian defined on the graph. Due to correlations between the $x_i$, the expectation of the product over neighbors of $i$ in Eq.\,\eqref{ni} does not factor in $j$, even on a tree. The standard simplifying feature that facilitates cavity analyses on trees or on locally tree-like systems is thus, at first sight, not available in the present case. Averaging Eqs.\,\eqref{ni} {\em is}, however, facilitated by the fact that --- {\em conditioned\/} on $x_i$ --- the averages over the $\big(\chi_{\{x_j\ge h\}} n_j^{(i)}\big)_{j\in \partial i}$ in Eqs.\,\eqref{ni} do factor in $j$, if the graph in question is a tree, and that such factorization becomes asymptotically exact on locally tree-like graphs in the thermodynamic limit. Analogous factorization is possible for averages over the $\big(\chi_{\{x_\ell\ge h\}} n_\ell^{(j)}\big)_{\ell \in \partial j\setminus i}$ in Eqs.\,\eqref{nji}, when conditioned on $x_j$.

Performing the average over the Gaussian field $\vx$ in this way, we obtain $g_i = \E_\vx[n_i] = \E_{x_i}\Big[\E_\vx[n_i|x_i]\Big]$ from Eq.\,\eqref{ni} as
\be
g_i\!\!=\!\!\E_{x_i}\Bigg[\chi_{\{x_i\ge h\}}\bigg(\!1\! - \!\prod_{j\in \partial i}\!\!
        \Big(1 - \E_\vx\big[\chi_{\{x_j\ge h\}} n_j^{(i)}\big| x_i\big]\Big)\bigg)\Bigg]
\ee
by factorization of conditional expectations, where $\E_{x_i}[\cdot]$ denotes an expectation evaluated over the single node marginal of $x_i$. Then, using further conditioning, we have 
\bea
\E_\vx\Big[\chi_{\{x_j\ge h\}} n_j^{(i)}\big| x_i\Big]\!\!
                    &=&\!\! \E_{x_j}\Big[\chi_{\{x_j\ge h\}} \big| x_i\Big]\nn\\
                    & &\!\! \times \E_\vx\Big[n_j^{(i)}\big| \{x_j\ge h\}, x_i\Big]\nn\\
                    &=&\!\! H_j(h|x_i)~g_j^{(i)}
\eea
where we have introduced 
\be
H_j(h|x_i) = \E_{x_j}\Big[\chi_{\{x_j\ge h\}}\Big| x_i\Big]
\label{Hji}
\ee
and 
\be 
g_j^{(i)} = \E_\vx\Big[n_j^{(i)}\big|\{x_j\ge h\}\Big]\ .
\label{gjidef}
\ee
In Eq.\,\eqref{gjidef} we have used the fact that for conditional expectations of observables pertaining to the cavity graph  such as $n_j^{(i)}$ we have $\E_\vx\big[n_j^{(i)}\big| \{x_j\ge h\}, x_i\big]=\E_\vx\big[n_j^{(i)}\big| \{x_j\ge h\}\big]$. Putting things together, we obtain
\be
g_i\!\! =\!\! \rho_i^h 
 \Bigg(\! 1\! -\!\E_{x_i}\bigg[\!\prod_{j\in \partial i}\!\! \Big(\! 1\! -\! H_j(h|x_i) g_j^{(i)}\Big)\Big|\{x_i\ge h\}\bigg]
\Bigg)
\label{gi}
\ee
with $\rho_i^h = \E_{x_i}[\chi_{\{x_i\ge h\}}]$. 

Following an entirely analogous line of reasoning and using the same sequence of conditionings, we can evaluate the $g_j^{(i)}$ defined in Eq.\,\eqref{gjidef} by evaluating the conditional average of $n_j^{(i)}$ using Eq.\,\eqref{nji}, giving
\be
g_j^{(i)}\! =\! 1 - \E_{x_j}\bigg[\!\prod_{\ell\in \partial j\setminus i}\!\!\Big(\! 1\!-\! H_\ell(h|x_j) g_\ell^{(j)}\Big)\Big|\{x_j\ge h\}\bigg]\ .
\label{gji}
\ee
Equations \eqref{gi} and \eqref{gji} for the $g_i$ and the $g_j^{(i)}$ can be evaluated, once single-site marginals and joint densities on adjacent sites of the Gaussian field defined by Eqs.\,\eqref{GFF} and \eqref{H} are known; the latter are required to evaluate the conditional  probabilities $H_j(h|x_i)$ defined in \eqref{Hji} (and similarly the $H_\ell(h|x_j)$ appearing in Eq.\,\eqref{gji}), while the former are needed to evaluate $x_i$ expectations in Eqs.\,\eqref{gi} and $x_j$ expectations in Eqs.\,\eqref{gji}, respectively. They are --- for large, heterogeneous locally tree-like random graphs --- most efficiently obtained by their own cavity type analysis, which has in fact been performed (directly in the thermodynamic limit) in \cite{Ku+07} for single-site marginals of harmonically coupled systems on random graphs, and in \cite{Ku08,Rog+08} in the context of the spectral problem of sparse symmetric random matrices. All that is needed are the (Gaussian) single-site marginals $P_i(x_i)$ of $P(\vx)$, as well as the corresponding single-site cavity marginals $P_j^{(i)}(x_j)$ for $j\in\partial i$ on the cavity graph $G^{(i)}$, in terms of which joint densities on adjacent sites are easily obtained. Key identities needed in the analysis are reproduced appendix \ref{secGauss} of this paper.  Single-site marginals and single-site cavity marginals are fully characterized by their inverse variances (or precisions) $\omega_i$  and $\omega_j^{(i)}$, respectively. The latter are obtained by solving the system \eqref{CavPrec} of cavity self-consistency equations in appendix \ref{secGauss}, while the former can be evaluated in terms of the $\omega_j^{(i)}$. The $H_j(h|x_i)$ and the $H_\ell(h|x_j)$ can be expressed in closed form in terms of error functions, but the conditional $x_i$-expectation of the product in Eq.\,\eqref{gi} and similarly the conditional $x_j$-expectation of the product in Eq.\,\eqref{gji} will have to be evaluated numerically.

With all ingredients thus available, Eqs.\,\eqref{gji} constitute a set of coupled self-consistency equation for the $g_j^{(i)}$. They can be solved iteratively at given level $h$ on large instances of locally tree-like (random) graphs, starting from random initial conditions. Using the solutions, one obtains the node-dependent percolation probabilities $g_i$ from Eqs.\,\eqref{gi}.\\

\subsection{Percolation Threshold and Near-Critcal Solution}

The value of the percolation threshold follows from a linear stability analysis of Eqs.\,\eqref{gji}. These equations {\em always\/} allow the trivial solution $g_j^{(i)} \equiv 0$. This solution becomes unstable, indicating the percolation transition, where the the largest eigenvalue of the Hessian of the r.h.s. of Eqs.\,\eqref{gji} evaluated at $g_j^{(i)} \equiv 0$ exceeds 1. The Hessian is a weighted version of a so-called non-backtracking matrix, with non-zero elements
\be
B_{(ij),(j\ell)} =  \E_{x_j}\big[H_\ell(h|x_j)\Big|\{x_j\ge h\}\big]
\label{nb}
\ee
for $j\in \partial i$ and $\ell\in\partial j\setminus i$, and $B_{(ij),(k\ell)} = 0$ otherwise. 

For $h\lesssim h_c$ it is expected, and indeed borne out by a numerical solution of the self-consistency equations Eqs.\,\eqref{gji} that the cavity probabilities $g_j^{(i)}$, hence the site-dependent percolation probabilities $g_i$ will be small. Performing an appropriately adapted weakly non-linear expansion of Eqs.\,\eqref{gji} as in \cite{KuRog17}, one obtains site dependent percolation probabilities to linear order in $h_c-h$ as
\be
g_i \simeq \alpha ~ (h_c-h) \sum_{j\in \partial i} v_j^{(i)}
\ee
where $\bm v = \big(v_j^{(i)}\big)$ is the Frobenius right eigenvector corresponding to the largest eigenvalue $\lambda_{\rm max}(B)\big|_{h=h_c}= 1$ of the non-backtracking matrix \eqref{nb} evaluated at $h_c$, normalized s.t. $||v||_1=1$, and $\alpha$ is an amplitude given in Eq.\,\eqref{alpha} of Appendix \ref{secAsympt}, which also includes a derivation of the $\cO((h_c-h)^2)$ contribution to the $g_i$.\\

\section{Thermodynamic Limit}
\label{secThDLim}
\subsection{Probabilistic Self-Consistency}
For random graphs in the configuration model class, i.e., the class of graphs that are maximally random subject to a given degree distribution $p_k= \mbox{Prob}(k_i=k)$, one can analyse the level-set percolation problem in the thermodynamic limit of infinite system size. We will first describe the analysis for homogeneous mass parameters $\mu_i\equiv \mu$ and thereafter introduce the modifications needed to capture node dependent mass parameters $\mu_i$.

Assuming that a limiting probability law for the joint distribution of the cavity precisions $\omega_j^{(i)}$ and the cavity probabilities $g_j^{(i)}$ exists, probabilistic self-consistency compatible with the self-consistency equations \eqref{CavPrec} for the $\omega_j^{(i)}$ and with Eqs.\,\eqref{gji} for the $g_j^{(i)}$ allows one to obtain the PDF $\tilde\pi(\tilde g,\tilde \omega)$ as follows. One averages the right hand sides of Eqs.\,\eqref{CavPrec} and \eqref{gji} over all realizations for which $\omega_j^{(i)} \in (\tilde \omega, \tilde \omega+\rd \tilde \omega]$ and  $g_j^{(i)} \in (\tilde g, \tilde g+\rd \tilde g]$ (see e.g. \cite{KuRog17} for a similar line of reasoning) to obtain 
\begin{widetext}
\be
\tilde\pi(\tilde g,\tilde \omega)\! =\! \sum_k \frac{k}{\langle k\rangle}p_k\! \int\! \Big[\prod_{\ell=1}^{k} \rd \tilde \pi(\tilde g_\ell,\tilde\omega_\ell) \rd\rho_K(K_\ell)\Big]\, \delta\Big(\tilde\omega\! -\!\Omega_{k-1}\Big) \delta\bigg(\tilde g\! -\! \Big(1\!- \!\E_{x}\Big[\prod_{\ell=1}^{k-1}\big(1\!-\! H_\ell(h|x)\,\tilde g_\ell\big)\big|\{x\ge h\}\Big]\Big) \bigg)
\label{tpi}
\ee
in which 
\be
\Omega_{q} = \Omega_{q}(\{\tilde\omega_\ell\}) = \mu + \sum_{\ell=1}^q \frac{K_\ell \tilde\omega_\ell}{K_\ell +\tilde\omega_\ell}\ ,
\ee
and
\be 
H_\ell(h|x) = \E_{x_\ell}\Big[\chi_{\{x_\ell\ge h\}}\Big|x \Big] =  H\bigg(\sqrt{K_\ell + \tilde \omega_\ell}\,\Big(h - \frac{K_{\ell} x}{K_{\ell} +\tilde\omega_\ell} \Big) \bigg)\ ,
\label{Hell}
\ee
is the probability that the multivariate Gaussian on the $\ell$-th node adjacent to a node with Gaussian component $x$ (to which it is coupled via $K_\ell$) does itself exceed the value $h$. In Eq.\eqref{tpi}, $\frac{k}{\langle k\rangle}p_k$  is the probability for a random neighbor of a node to have degree $k$, and the expectation w.r.t. $x$ is evaluated for $x \sim \cN(0,1/\Omega_k)$, while $\rho_K(\cdot)$  is the PDF of the $K_\ell$. Moreover, we have introduced the shorthand $\rd \tilde \pi(\tilde g_\ell,\tilde\omega_\ell) = \rd \tilde g_\ell \rd \tilde\omega_\ell\, \tilde \pi(\tilde g_\ell,\tilde\omega_\ell)$, and similarly $\rd \rho_K (K_\ell)= \rd K_\ell \rho_K(K_\ell)$. Equation \eqref{tpi} is very efficiently solved by a population dynamics algorithm. From the solution we obtain
\bea 
\pi(g,\omega)\!\! &=&\!\!\sum_k\! p_k\!\int\!\! \Big[\prod_{\ell=1}^k \rd \tilde \pi(\tilde g_\ell,\tilde\omega_\ell)\rd\rho_K(K_\ell)\Big]\,\delta\Big(\omega\! -\!\Omega_{k}\Big) \delta\Big(g\!-\!\rho^h \Big(1- \E_{x}\Big[\prod_{\ell=1}^{k}\big(1\!-\! H_\ell(h|x)\,\tilde g_\ell\big)\big|\{x\ge h\}\Big] \Big)\ .
\label{pi}
\eea 
for the limiting joint distribution of single-site percolation probabilities $g_i$ and single site precisions $\omega_i$. In this equation $\rho^h  = \E_{x}[\chi_{x \ge h}]$, and we once more have $x \sim \cN(0,1/\Omega_k)$.

\end{widetext}

If mass parameters $\mu_i$ are indeed heterogeneous, one starts from the assumption that a limiting law $\rd \tilde\pi(\tilde\omega,\tilde g,\mu) =  \rd \tilde\pi(\tilde\omega,\tilde g|\mu)\rd\rho(\mu)$ exists, with $\rho(\cdot)$ the PDF of the locally varying mass parameters. One then obtains the joint PDF $\tilde\pi(\tilde\omega,\tilde g,\mu)$ in the same manner as before in the simpler version with homogeneous mass parameters, giving a modified probabilistic self-consistency equation of the form
\begin{widetext}
  \bea
  \tilde\pi(\tilde g,\tilde \omega,\mu) &=& \rho(\mu) \sum_k \frac{k}{\langle k\rangle}p_k \int \Big[\prod_{\ell=1}^{k} \rd \tilde \pi(\tilde g_\ell,\tilde\omega_\ell,\mu_\ell) \rd\rho_K(K_\ell)\Big]\, \delta\Big(\tilde\omega -\Omega_{k-1}\Big)\nn\\
  & & \hspace{30mm}\times \delta\bigg(\tilde g - \Big(1- \E_{x}\Big[\prod_{\ell=1}^{k-1}\big(1- H_\ell(h|x)\,\tilde g_\ell\big)\big|\{x\ge h\}\Big]\Big) \bigg)
  \label{tpi-mu}
  \eea 
Thus, it turns out that the conditional PDF $\tilde\pi(\tilde\omega,\tilde g|\mu)$ is formally expressed by the r.h.s. of the simpler self-consistency equation \eqref{tpi} for homogeneous mass parameters, if the $\tilde \pi(\tilde g_\ell,\tilde\omega_\ell)$ in Eq.\,\eqref{tpi} are interpreted as the $(\tilde\omega_\ell,\tilde g_\ell)$-marginals of the $\tilde\pi(\tilde\omega_\ell,\tilde g_\ell,\mu_\ell)$.  From the solution of Eq.\,\eqref{tpi-mu} we obtain
\bea 
\pi(g,\omega,\mu)\!\! &=&\!\!\rho(\mu) \sum_k\! p_k\!\int\!\! \Big[\prod_{\ell=1}^k \rd \tilde \pi(\tilde g_\ell,\tilde\omega_\ell,\mu_\ell)\rd\rho_K(K_\ell)\Big]\,\delta\Big(\omega\! -\!\Omega_{k}\Big)\nn\\ 
& & \times \delta\Big(g\!-\!\rho^h \Big(1- \E_{x}\Big[\prod_{\ell=1}^{k}\big(1\!-\! H_\ell(h|x)\,\tilde g_\ell\big)\big|\{x\ge h\}\Big] \Big)
\label{pi-mu}
\eea
as the joint pdf of local percolation probabilities, precisions and mass parameters.
\end{widetext}

\subsection{Random Regular Graphs}
Specializing to random regular graphs (RRGs) with uniform couplings, more explicit results can be obtained. The key observation is that in the thermodynamic limit all nodes and all edges of the system are equivalent. Hence the self-consistency equation for the uniform cavity precisions on a RRG of degree $c$ (or $c$RRG) reads
\be
\tilde\omega = \mu + (c-1) \frac{K\tilde\omega}{K+ \tilde\omega}\ .
\label{tomRRG}
\ee
This equation is solved by
\be
\tilde\omega_\pm\! =\!\frac{1}{2}\Big[\mu+ K(\!c-\!2) \pm \sqrt{[\mu\!+\!K(\!c-\!2)]^2\! +\! 4K\mu}\Big]\ ,      
\label{tomcRRG}
\ee
the relevant (physical) solution being $\tilde\omega=\tilde\omega_+$. This entails a self-consistency equation for the uniform cavity percolation probabilities $g_j^{(i)}\equiv \tilde g$ of the form
\be
\tilde g = 1 - \E_{x}\Big[\Big(1 - H(h|x)\,\tilde g\Big)^{c-1}\Big|\{x\ge h\}\Big]\ ,
\label{tg}
\ee
in which $H(h|x)$ is a conditional expectation of the type defined in Eq.\,\eqref{Hji}, evaluated on the $c$RRG, and $x\sim \cN(0,1/\omega)$, with
\be
\omega = \mu + c \frac{K\tilde\omega}{K+ \tilde\omega} = \tilde\omega + \frac{K\tilde\omega}{K+ \tilde\omega}
\label{omcRRG}
\ee
the uniform single-site precision on the $c$RRG. Equation \eqref{tg} is a simple scalar equation for $\tilde g$ which is easily solved numerically. It always has the trivial solution $\tilde g =0$, which becomes unstable below a critical value $h_c$ of the level $h$ wich follows from a linear stability analysis of Eq.\,\eqref{tg} and is given as the solution of
\be
(c-1) \E_{x}\Big[H(h|x)\Big|\{x\ge h\} \Big] = 1\ .
\label{hc}
\ee
This equation has to be solved numerically. In the limit of large $c$, however, it is possible to provide an asymptotic expansion for the solution $h_c$ in closed form.  It is performed in Appendix \ref{secAsymptRRG}, both for the case $K = K(c) = 1$ for $K = K(c) = 1/c$.

For $K = K(c) = 1$ we obtain
\be
h_c \simeq \sqrt{\frac{2}{\mu+c-1}[y_0 + o(y_0)]} 
\ee
with $y_0 = \ln\Big(\frac{c-1}{2\sqrt{\pi}}\Big)$ defined in Eq.\,\eqref{defy0}. Thus $h_c \searrow 0$ as $c\to \infty$ in this case.

For $K = K(c) = 1/c$, on the other hand, the result is
\be
h_c \simeq \sqrt{\frac{2}{\mu+(c-1)/c}[y_0 + o(y_0)]} 
\ee
hence $h_c$ diverges as $c\to \infty$.

From the solution of Eq.\,\eqref{tg} at $h< h_c$ one obtains
\be
g = \rho^h \Big(1 - \E_{x}\Big[\Big(1 - H(h|x)\,\tilde g\Big)^{c}\Big|\{x\ge h\} \Big]\Big)\ ,
\ee
with $\rho^h = \E_{x}[\chi_{\{x\ge h\}}]$, as the value of the percolation probability $g$ at level $h < h_c$. 

\section{Results}
\label{secRes}
In what follows we present results obtained for the Gaussian level-set percolation problem for random graphs in the configuration model class. Given the huge number of possible variations of parameters and parameter combinations that could be contemplated, the results presented below can of course only illustrate general trends and not be exhaustive. We begin by presenting results for systems uniform mass parameters $\mu_i\equiv \mu$ and homogeneous edge weights $K_{ij}\equiv 1$ and thereafter briefly illustrate the effect of heterogeneous mass parameters and edge weights.

\begin{figure}[h!]
  \includegraphics[width = 0.45\textwidth]{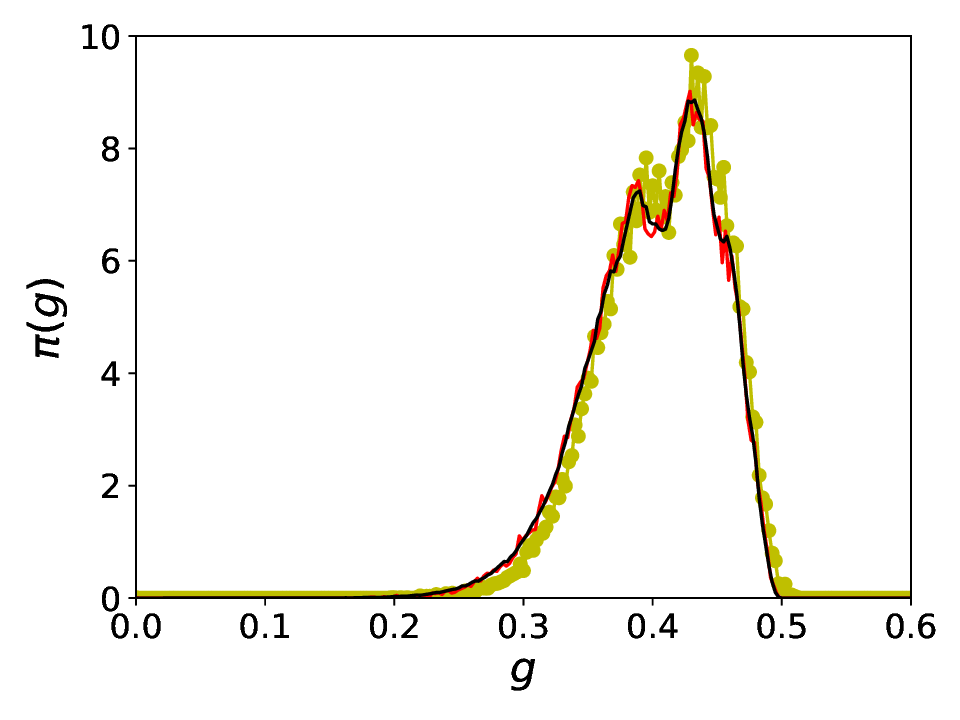}\\
  \caption{(color online) Distribution $\pi(g)$ or local percolation probabilities for a shifted Poisson degree distribution $k \sim 2 +\mbox{Poiss}(1)$ at $h=0$ and $\mu=0.1$. We compare {\bf (i)} results of a numerical simulation of a single instance of a graph of $N=50,000$ vertices, averaging over 5,000 realizations of Gaussian field configurations to obtain the PDF of the $g_i$ (yellow dots), with {\bf (ii)} results of a single instance cavity analysis for the same graph (red solid line), and {\bf (iii)} the result of an analysis in the thermodynamic limit (black solid line).}
  \label{FigSCPER3m2h0mu0p1}
\end{figure}

\subsection{Testing the Theory}

In Fig.\,\ref{FigSCPER3m2h0mu0p1}, we present a distribution of level-set percolation probabilities for a system with a shifted Poisson degree distribution with $k_{\rm min} = 2$ and $\langle k\rangle = 3$, which demonstrates that the theoretical analyses agree very well with a numerical simulation. Simulations are, of course, affected by finite size effects (creating details depending on the specific single realization of the generated random graph) {\em and\/} by sampling fluctuations (that are created by approximating average local percolation probabilities using empirical averages over a finite sample of $N_s = 5,000$ realizations of the multi-variate Gaussian, which are themselves generated through a Metropolis Monte-Carlo procedure), while the single instance cavity analysis is only affected by finite size effects. 

\begin{figure}[h!!]
\includegraphics[width = 0.45\textwidth]{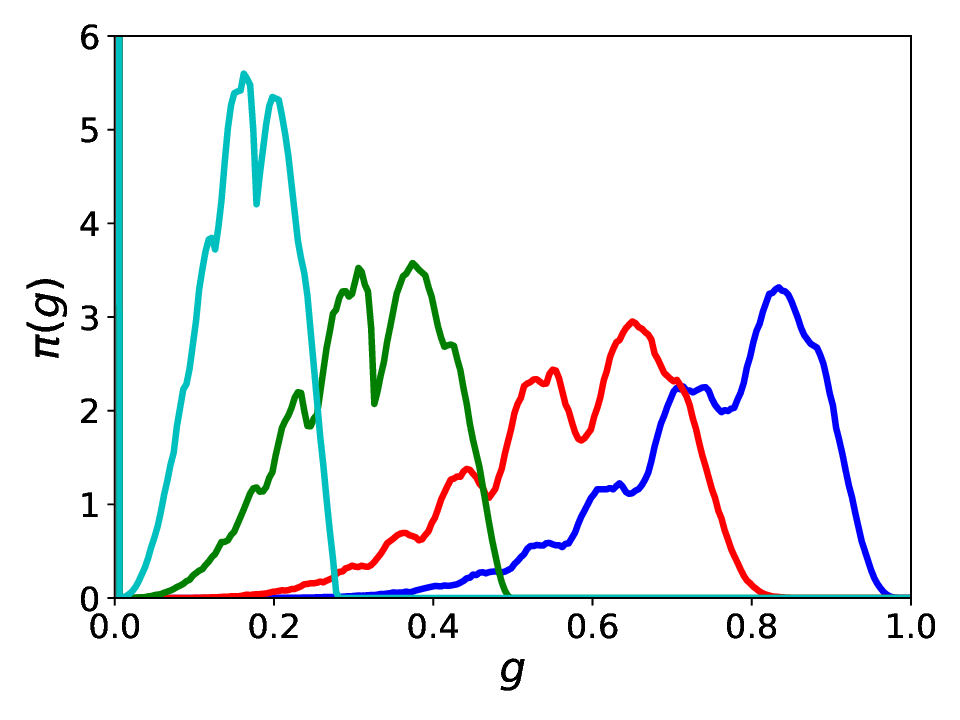}\\ 
\includegraphics[width = 0.45\textwidth]{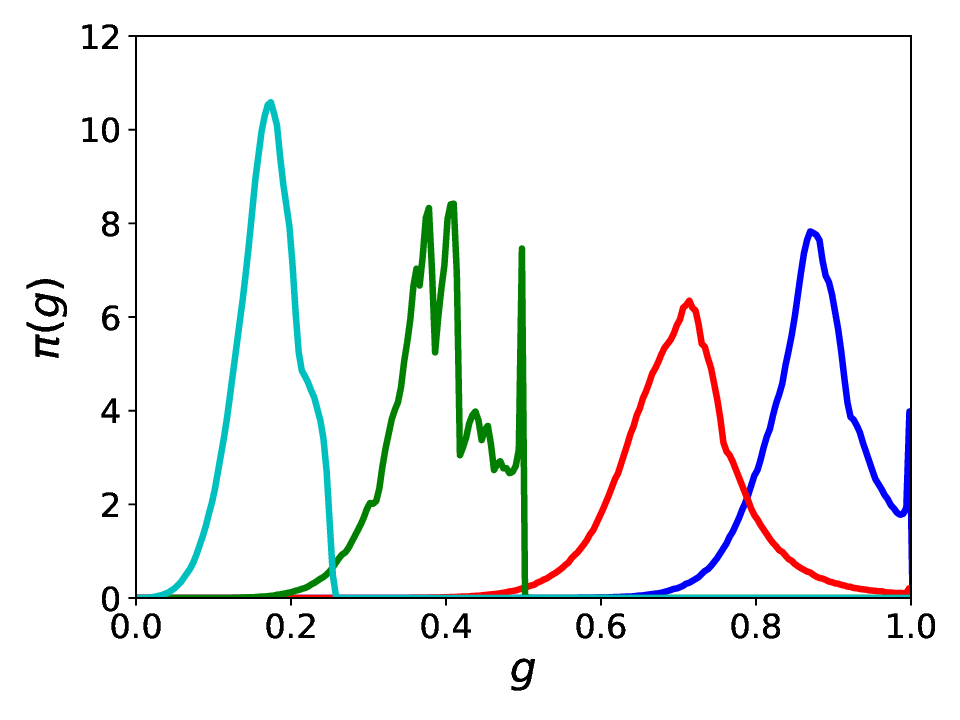}
\caption{(color online)  Top panel: Distribution $\pi(g)$ of local percolation probabilities at levels $h=-1$ (blue), $h=-0.5$ (red), $h=0$ (green), and $h=0.25$ (cyan)  from right to left for an ER graph with mean degree 2, evaluated in the thermodynamic limit. Bottom panel: distributions for the same values of $h$ are displayed for a graph with power law degree distribution, $p_k \sim k^{-3}$ for $2\le k \le 125$.} 
  \label{FigER3pw3m2}
\end{figure}

\subsection{PDFs of Local Percolation Probabilities}

Figure \ref{FigER3pw3m2} shows the distribution $\pi(g)$ of local percolation probabilities for a massless Gaussian field at levels $h=-1$, $h=-0.5$, $h = 0$ and $h=0.25$ on an Erd\H{o}s-R\'enyi (ER) graph of mean degree $\langle k\rangle = 2$ and for a system with power law degree distribution $p_k \sim k^{-3}$ for $2\le k \le 125$ in the thermodynamic limit. In the ER case, the original graph contains finite components, generating a $\delta$-peak at 0 in $\pi(g)$, whereas in the latter it doesn't. In both cases, the center of mass, i.e. the average percolation probability of the distributions decreases with increasing value of $h$ as expected, with $\langle g\rangle \simeq 0.60, 0.46, 0.25$ and 0.13 at $h = -1.0, -0.5, 0$, and $h=0.25$, respectively for the ER2 graph, and $\langle g\rangle \simeq 0.87, 0.70, 0.39$, and 0.17 at these $h$ values for the system with power-law degree distribution. However, the shape of the distributions also changes markedly with the level $h$, thus carrying information that goes far beyond the respective average percolation probabilities. 

\begin{figure}[h!]
\includegraphics[width = 0.45\textwidth]{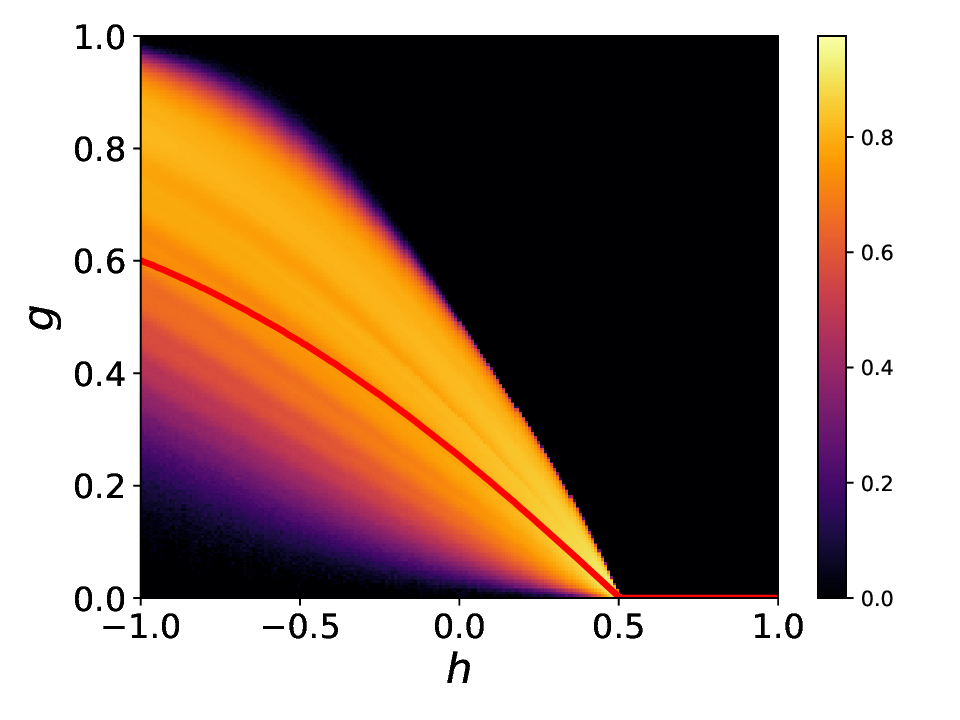}\\ 
\includegraphics[width = 0.45\textwidth]{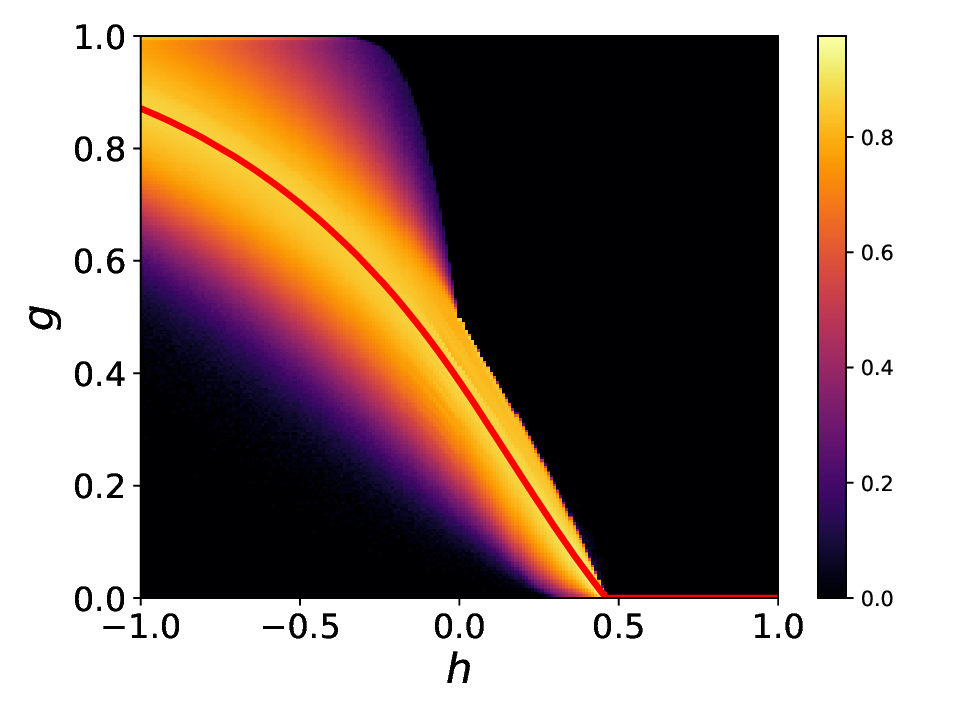}
\caption{(color online)  Top panel: Heat-map of the distribution $\pi(g)$ of local percolation probabilities for massless Gaussian field on an ER graph with mean degree 2, evaluated in the thermodynamic limit. Bottom panel: heat-map for a massless Gaussian field on a graph with power law degree distribution, $p_k \sim k^{-3}$ for $2\le k \le 125$. Values of $\pi(g)$ are non-linearly transformed according to $\pi(\cdot) \to \sqrt{\pi(\cdot)}/(0.4 + \sqrt{\pi(\cdot)})$ because of their large dynamical range, in order to achieve a good resolution also at low values of $\pi(g)$. Mean percolation probabilities $\langle g\rangle>$ as functions of the level $h$ are for both cases shown as solidl red lines.} 
  \label{FighHeatmaps}
\end{figure}

In order to provide a a more global --- albeit qualitative --- view of level set percolation, heat-maps of distributions of local percolation probabilities together with mean percolation probabilities for a range of values of the level $h$ are displayed for the same systems in Fig.\,\ref{FighHeatmaps}. The critical level $h_c$ is $h_c\simeq 0.52$ for massless Gaussians defined on the ER-2 graph, and $h_c \simeq 0.47$ for massless Gaussians on a graph with power law degree distribution, $p_k \sim k^{-3}$ for $2\le k \le 125$.

\subsection{Correlations Between Marginal Variances and Local Percolation Probabilities}

\begin{figure}[t!]
  \includegraphics[width = 0.45\textwidth]{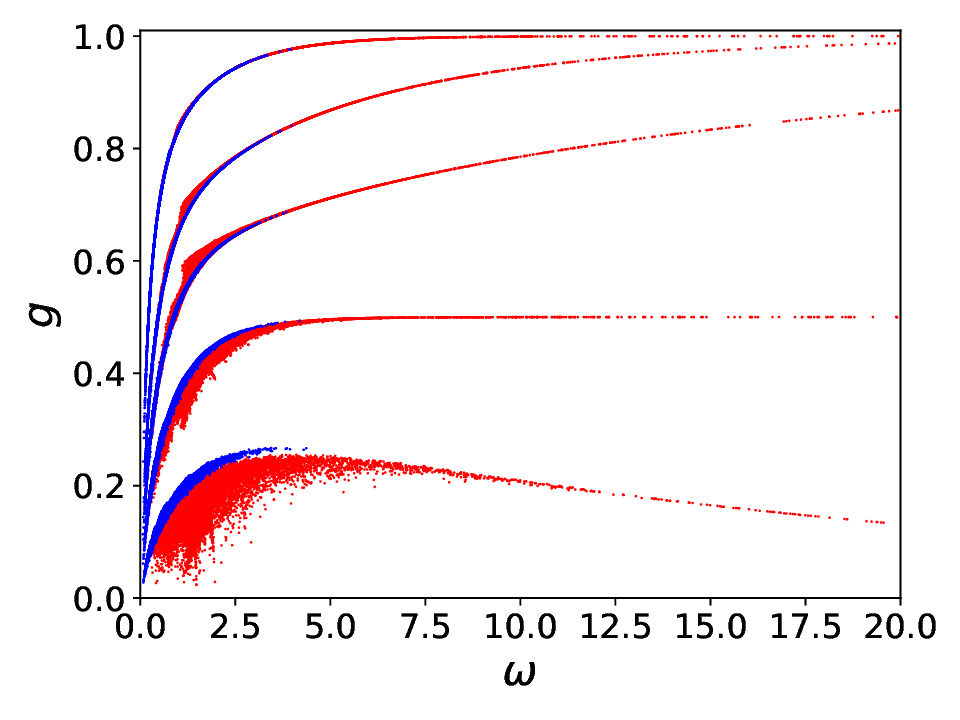}\\
  \includegraphics[width = 0.45\textwidth]{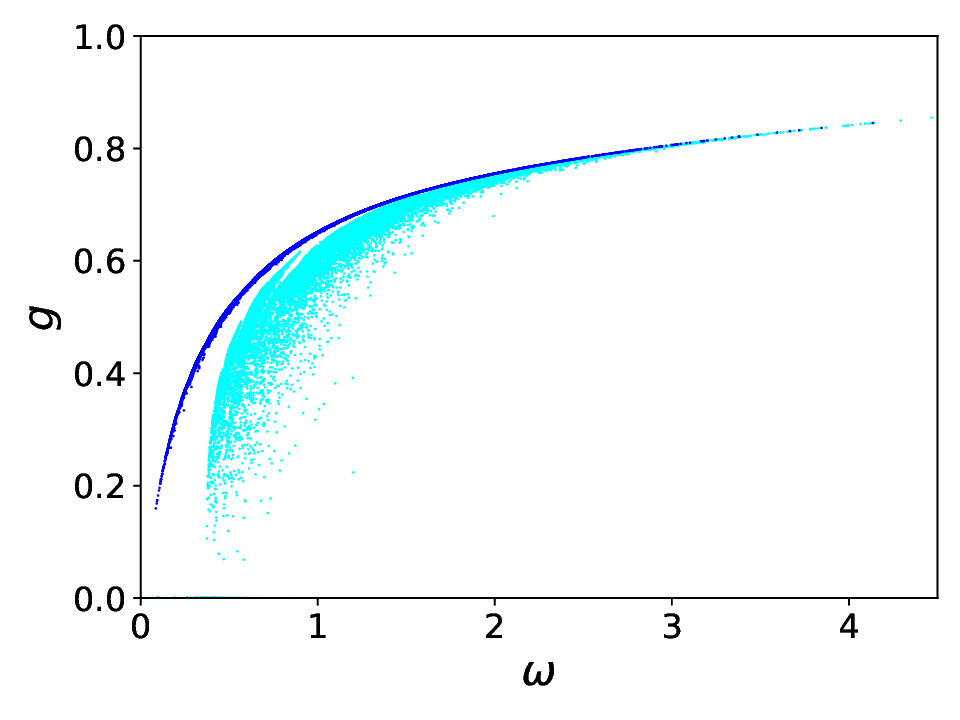}
\caption{(color online) Top panel: Scatter-plot of local level set percolation probabilities $g_i$ at levels $h=-1.0$, $h=-0.5$, $h=-0.25$, $h=0$ and $h=0.25$ (top to bottom) against local single node precisions $\omega_i$ for a massless Gaussian field defined on an ER graph  of size $N=100,000$ with mean degree 2 (blue dots), and for a a graph of the same size with power law degree distribution $p_k \sim k^{-3}$ for $2\le k \le 125$ (red dots). Bottom panel: For $h=-0.5$ the data for the massless Gaussian field on the ER graph of mean degree 2 (blue dots) are displayed together with data obtained for a system of the same type, but now for a Gaussian field with mass parameter $\mu=0.1$ (cyan dots).}
\label{FigOmgScatt}
\end{figure}

In Fig.\ref{FigOmgScatt} we display scatter-plots of marginal level-set percolation probabilities $g_i$ versus marginal Gaussian precisions $\omega_i$ at various values for the level $h$, again for an ER graph with mean degree 2 (blue dots), and for a a graph of the same size with power law degree distribution $p_k \sim k^{-3}$ for $2\le k \le 125$ (red dots). The results suggest that there is a remarkably strong correlation between the $g_i$ and the $\omega_i$ values, which is almost perfect for negative $h$ values but weakens (at small $\omega_i$ values) for the system with power law degree distribution as $h\nearrow 0$. The stong correlation between the two at negative $h$ appears to be even insensitive to the underlying graph structure, with the curves for the graph with power-law distributed degrees overlapping with those for the ER graph, but extending them to larger $\omega$ and $g$ values. The second panel shows that the nearly perfect correlation between the $g_i$ and the $\omega_i$ for the Erd\H{o}s-R\'enyi graph is also weakened if the Gaussian field acquires a non-zero mass $\mu >0$.

\subsection{The Effect of Randomly Varying Mass Parameters and Random Couplings}
Results so far reported were obtained for systems with homogeneous mass parameters $\mu_i\equiv \mu$ and uniform couplings $K_{ij}\equiv K = 1$. In what follows we briefly look at the effect of introducing randomly varying mass parameters $\mu_i$ and randomly varying edge weights $K_{ij}$.

\begin{figure}[h!]
  \includegraphics[width = 0.45\textwidth]{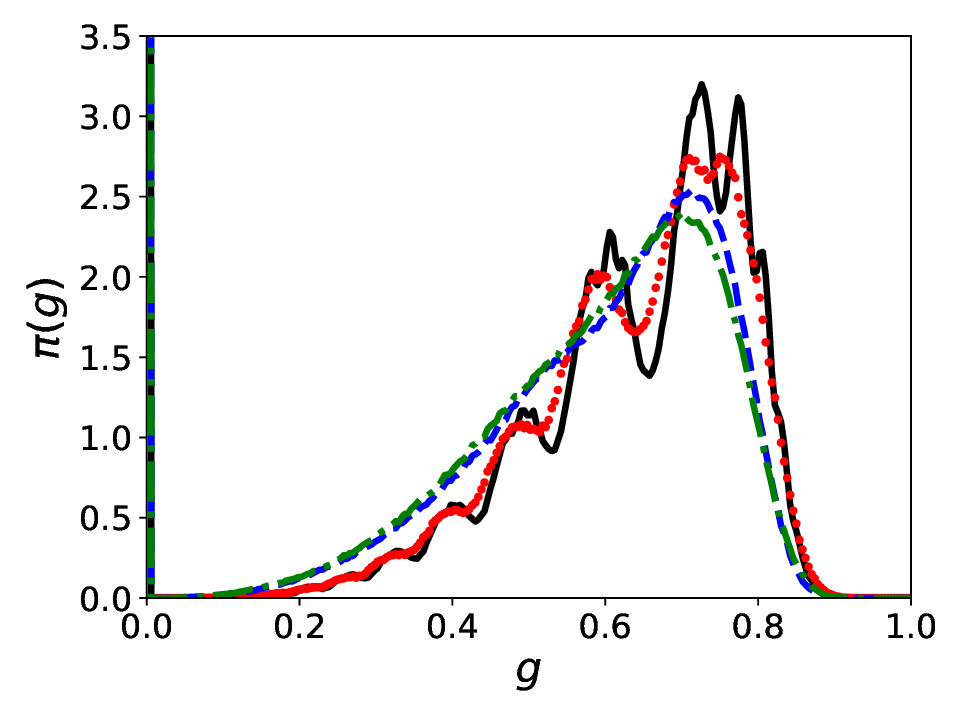}
  \caption{\!\!(color online) Distribution $\pi(g)$ of local percolation probabilities for an ER-2 graph at level $h=-0.5$ for a system (i) with homogeneous mass parameter $\mu=0.5$ and homogeneous edge parameters $K_{ij}\equiv 1$ (black solid curve), (ii) a system with exponentially distributed $\mu_i$ of mean  0.5, but homogeneous edge parameters $K_{ij}\equiv 1$ (red dotted curve), (iii) a system with $\mu_i\equiv 0.5$ and  exponentially distributed edge weights $K_{ij}$ of mean 1 (blue dashed curve), and (iv) a system where both the $\mu_i$ and the $K_{ij}$ are exponentially distributed with means 0.5 and 1 (green dot-dashed curve), respectively.}
\label{Figmuvar0p5}
\end{figure}

Replacing a homogeneous mass parameter $\mu$ by exponentially distributed mass parameters $\mu_i$ of the same mean has a barely detectable effect if $\mu$ is small compared to the value of the (uniform) coupling $K_{ij} \equiv 1$. E.g. (not shown), for $\mu=0.1$ the effect of introducing exponentially varying masses of the same mean is barely detectable on a graph of the PDF of locally varying percolation probabilities. However, as shown in Fig.\,\ref{Figmuvar0p5} for a larger values of $\mu = 0.5$, replacing a homogeneous mass parameter by  exponentially distributed mass parameters of the same mean does have a clearly notable smoothing effect on $\pi(g)$. Introducing heterogeneity of the edge weights, by replacing them with exponentially distributed weights of the same mean 1 has a significantly stronger smoothing effect, while adding heterogeneity of the $\mu_i$ to the heterogeneity of the edge weights does not have a very strong additional smoothing effect.

\subsection{Results for Random Regular Graphs}

\begin{figure}[h]
    \includegraphics[width = 0.45\textwidth]{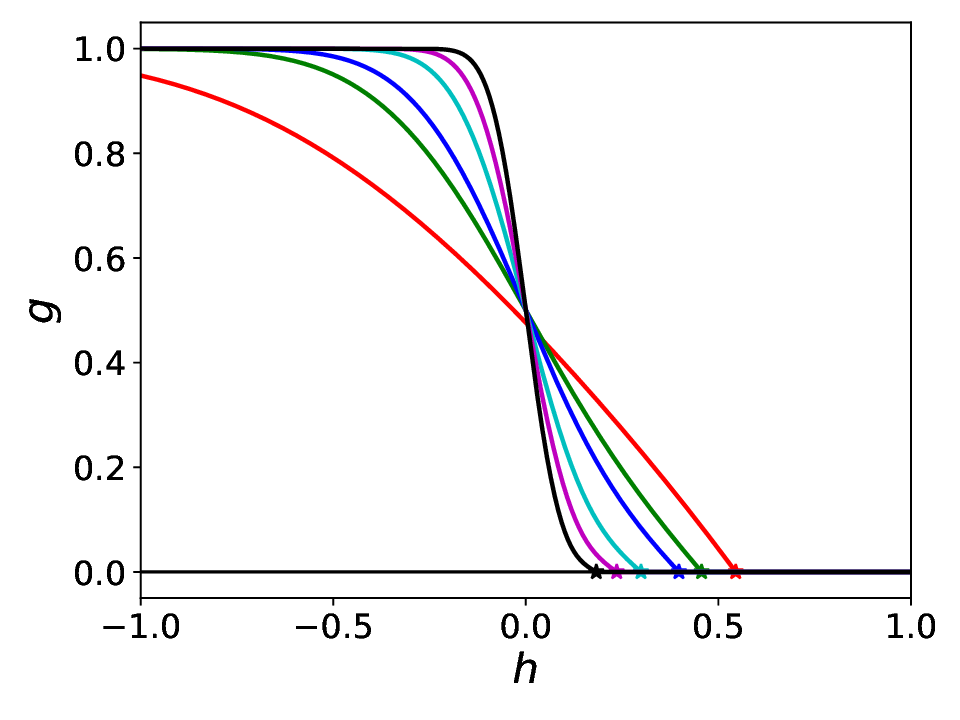}
    \caption{(color online) Percolation probability $g$ as a function of the level $h$ for $c$RRGs with $\mu=0$, $K=K(c)\equiv 1$ and $c = 4$, 12, 20, 50, 100, and 200 (red, green, blue, cyan, magenta and black curves, respectively). The steepness of the curves increases with $c$. Critical levels $h_c$ as obtained from Eq.\,\eqref{hc} are indicated as asterisks. For the 6 values of $c$ shown here, they decrease with increasing $c$, and they agree perfectly with results of a numerical solution of Eq.\,\eqref{tg}.}
    \label{FigRRG4-200}
\end{figure}

\begin{figure}[h!]
  \includegraphics[width = 0.45\textwidth]{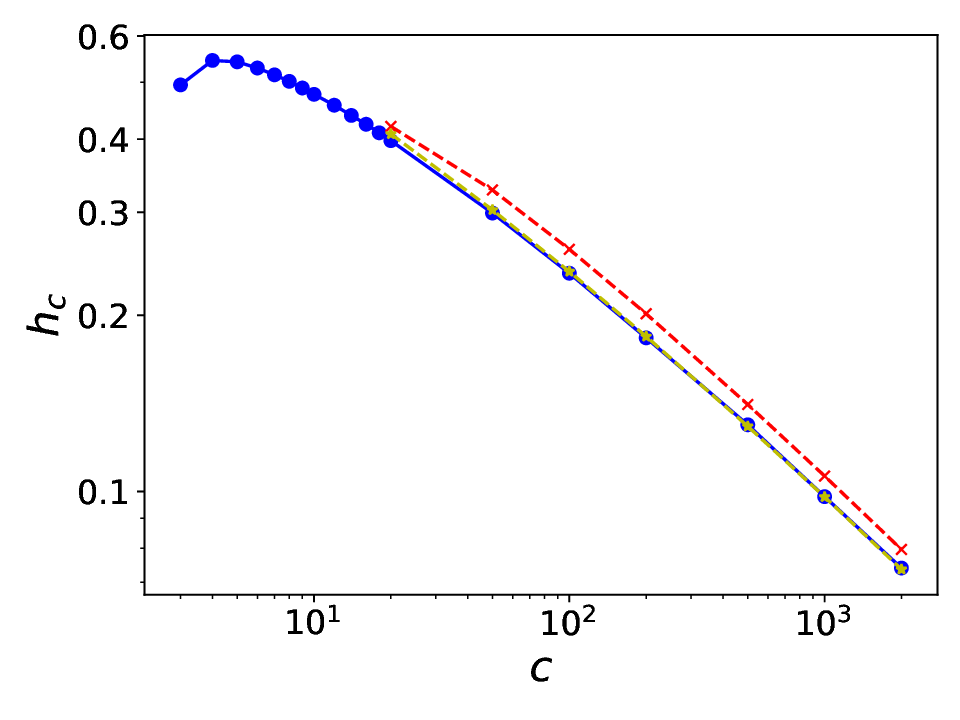}
  \caption{\!\!(color online) Degree dependent critical thresholds $h_c$ for $c$RRGs with $\mu=0$, $K=K(c)\equiv 1$ and $c\in\{3,4,5,6,7,8,9,10,12,14,16,18,20,50,100,200,500,$ $1000,2000\}$ (blue dots), with solid blue line as a guide to the eye. For the larger degrees $c\ge 20$ in this list, results of the asymptotic analysis of Eq.\,\eqref{hc} are shown as well: red crosses show $h_{c0}$, i.e., the lowest order of the asymptotic expansion of $h_c$ described by Eqs.\,\eqref{yna}, \eqref{hna} and while yellow asterisks represent $h_{c1}$ which includes the first sub-dominant corrections (with dashed red and yellow lines as guides to the eye.)}
  \label{Fig-hc-cRRG}
\end{figure}

In random regular graphs, all nodes and all edges are equivalent, and as a result there are no local variations of percolation probabilities. We show percolation probabilities as functions of the level $h$ for $c$RRGs with uniform couplings $K=1$ and $\mu=0$ for six different values of $c\ge 4$ in Fig.\,\ref{FigRRG4-200}. For the range of $c$ values shown critical percolation thresholds $h_c$ are decreasing with increasing $c$, but, as shown in Fig.\,\ref{Fig-hc-cRRG} there is non-monotonicity of $h_c$ as a function of $c$ in the range $c\in\{3,4,5\}$. 

\begin{figure}[t!]
  \includegraphics[width = 0.45\textwidth]{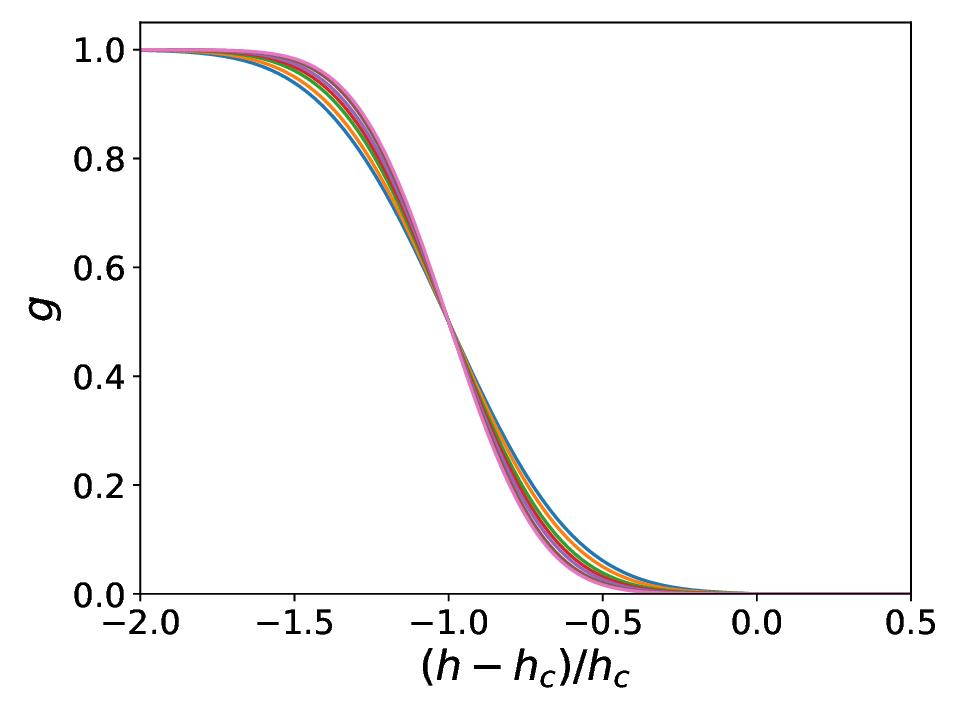}
  \caption{\!\!(color online) Percolation probability $g$ as a function of the scaling variable $(h-h_c)/h_c$ for $c$RRGs with $\mu=0$, $K=K(c)\equiv 1$ and $c = 1000$, 2000, 5000, 10000, 20000, 50000, and 100000, with the steepness of the curves increasing with $c$.}
  \label{Fig-cRRG-scaled}
\end{figure}

Results for $c\ge 20$ agree almost perfectly with the findings of the asymptotic analysis of Eq.\,\eqref{hc} performed in Appendix \ref{secAsymptRRG}. They show that $h_c\searrow 0$, as $c\to \infty$. At first sight the results of Fig.\,\ref{FigRRG4-200} suggest that the percolation transition itself becomes discontinuous in the large-$c$ limit. Upon closer inspection, however, this is seen to be an artifact of the fact that $h_c\searrow 0$, as $c\to \infty$. Plotting percolation probabilities as a function of the proper scaling variable $(h-h_c)/h_c$ reveals that the transition remains continuous at large $c$ as shown in Fig.\,\ref{Fig-cRRG-scaled}. In that Figure the point $(h-h_c)/h_c= -1$ corresponds to $h=0$. Results indicate the emergence of a limiting scaling function as $c\to\infty$. It turns out (not shown) that the corresponding scaled plots for the case $K=1/c$ are for the $c$ values presented here virtually indistinguishable from those of the case $K(c)\equiv 1$ presented in  Fig.\,\ref{Fig-cRRG-scaled}, although in that case the critcal percolation threshold diverges, whereas it converges to 0 for the the case with $K(c)\equiv 1$ shown here.

\section{Summary and Discussion}
\label{secSummary}
In the present paper we presented a full solution of the problem of level-set percolation of Gaussian free fields on locally tree-like random graphs. Our solution is based on a cavity or message passing approach and can be evaluated both for finite large instances and in the thermodynamic limit of infinite system size for random graphs in the configuration model class with finite mean degree. Finite single instance solutions require the simultaneous solution of a set self-consistency equations for locally varying single-node cavity percolation probabilities $g_j^{(i)}$ {\em and\/} for the locally varying single node cavity precisions $\omega_j^{(i)}$, with solutions of the former depending on solutions of the latter. The solution in the thermodynamic limit instead requires solving a non-linear integral equation for their joint PDF $\tilde\pi(\tilde g,\tilde \omega)$ derived form a probabilistic self-consistency arguments (in case of homogeneous mass parameters $\mu_i\equiv\mu$) and for the joint PDF $\tilde\pi(\tilde g,\tilde \omega,\mu)$ (in case of locally varying mass parameters). We found our results to be in excellent agreement with simulations. 

It is worth pointing out that the probabilistic self-consistency conditions describing the Gaussian level-set percolation problem do {\em not\/} allow one to extract a self-consistency condition for the {\em average\/} cavity percolation probability $\langle \tilde g\rangle$ and thereby the average $\langle g\rangle$ of the locally varying percolation probability, in marked contrast to the case of independent Bernoulli percolation on random graphs. In the present case, therefore, analytic control over locally varying (cavity) percolation probabilities is {\em essential\/} for the analysis of the problem, whereas in the case of independent Bernoulli percolation it could be seen as merely providing a more detailed level of analysis.

Distributions of level dependent local percolation probabilities were obtained both for Erd\H{o}s-R\'enyi graphs and for graphs with a fat tailed degree distribution described by a power-law, and they were found to exhibit a considerable amount of structure. We believe that this structure can to some extent be rationalized in terms of distributions of local environments of nodes in a manner analogous to the much simpler case of independent Bernoulli percolation \cite{KuRog17}. We note, however, that the {\em shape\/} of these distributions markedly changes with the level $h$, entailing that the relation between their structure and the distribution of local environments is less direct than in the case of Bernoulli percolation. We therefore leave a closer investigation of this issue to future publications. Not unexpectedly, some of the structure exhibited by the distributions of local percolation probabilities disappears upon introducing locally varying mass parameters $\{\mu_i\}$ or heterogeneous couplings $\{K_{ij}\}$. 

A remarkable and unexpected finding concerns the very strong correlation between marginal precisions $\omega_i$ and level-dependent local percolation probabilities $g_i$ in the case of a massless Gaussian field, which for negative values of the level $h$  is almost perfect and appears to be insensitive even to the underlying graph structure. However, this almost perfect correlation is weakened for networks with power-law distributed degrees, if the level $h$ at which the $g_i$ are evaluated approaches zero, or in general, if the Gaussian field acquires a non-zero mass. 

Simplifications in the analysis are possible in the case of $c$RRGs for which the level dependent uniform single-node cavity percolation probability $g_j^{(i)}\equiv \tilde g$ is obtained as solution of a single scalar equation, from which in turn the uniform percolation probability $g$ is easily evaluated. Asymptotic analysis reveals that for $K=K(c)=1$ the critical percolation threshold $h_c$ approaches 0 in the large $c$ limt, while $h_c$ diverges in this limit if edge-weights are scaled as $K=K(c)=1/c$.

While our methods are non-rigorous, they are expected to be exact in the large system limit. Indeed, in cases where this could be investigated, self-consistency equations obtained using the decorrelation assumption of the cavity approach for locally tree-like systems in the large system limit agree with those obtained using a saddle point approximation for the evaluation of a partition function known to be exact in the thermodynamic limit \cite{Ku+07, Ku08, Rog+08}.

Some of the methods and heuristics used in the present paper should be useful for the analysis of wider classes of level-set percolation problems. For instance, generalizing our methods to continuous multivariate {\em non}-Gaussian distributions described, e.g., in terms of Gibbs distributions with pair interactions including anharmonicities is in principle in reach of our methods. However, the analysis will be {\em significantly\/} more complex, as it requires to replace the self-consistency equations for single node cavity precisions by self-consistency equations for {\em entire functions\/}, viz. effective single-node cavity potentials (see \cite{Ku+07}) in the case of a large finite instance analysis, while it requires to solve a non-linear integral equation for the self-consistent distribution of these functions in the thermodynamic limit. On the other hand, it should be relatively straightforward to reformulate our methods to analyse, e.g., level-set percolation {\em of local fields\/} for disordered Ising models defined on random graphs, and thereby to level-set percolation for single-node magnetizations.  

Beyond generalizations of this type, it would also be interesting to investigate, whether the approach of \cite{KuRog17} which is capable of giving distributions of the sizes of {\em finite\/} clusters, both in the non-percolating and in the percolating phase, can be carried over to the present case of (Gaussian) level-set percolation. Another as yet unsolved problem concerns the stability analysis of the integral equation \eqref{tpi}, which could in principle allow one to obtain critical percolation levels $h_c$ for configuration model networks directly in the thermodynamic limit. 

In the context of statistical inference, level-set percolation could be used to assess the damage due to some form of contaminant spreading through a network, in cases where actual damage only occurs above a certain concentration of the contaminant.

We hope to address some of these problems in the near future.

\appendix
\section{Gaussian Identities}
\label{secGauss}
Averages and conditional averages of indicator functions as they appear in the theory require evaluating expectations over node-dependent single-site marginals of the multivariate Gaussian defined by Eqs.\,\eqref{GFF}, \eqref{H}, as well as averages over conditional distributions of these Gaussians, when conditioned w.r.t. the value of the Gaussian on a neighboring vertex.
 
Their evaluation uses cavity type reasoning of the same form used to compute single-node marginals and self-consistency equations for single-node cavity marginals as used before in the context of the theory of harmonically coupled systems on graphs \cite{Ku+07} or the theory of sparse random matrix spectra \cite{Ku08,Rog+08}. We collect key identities here.

For a free multivariate Gaussian field on a graph with joint Gaussian density given by Eqs.\,\eqref{GFF} and \eqref{H}, all marginals are themselves Gaussian, and of the form
\be
P_i(x_i) = \frac{1}{Z_i} \exp\Big(-\frac{1}{2} \omega_i x_i^2\Big)  
\label{GMarg}
\ee
with $Z_i = \sqrt{2\pi/\omega_i}$ and $\omega_i$ denoting their precisions (inverse variances). On a tree, and approximately on a locally tree-like graph, we have
\bea
P_i(x_i)\!&\propto&\!\re^{-\frac{1}{2} \mu_i x_i^2}\nn\\
& &\!\!\!\!\!\times\!\!\prod_{j\in\partial i}\!\int \rd x_j\,\re^{-\frac{1}{2} K_{ij}(x_i-x_j)^2} P_j^{(i)}(x_j)\ ,
\label{MargCav}
\eea
in which the $P_j^{(i)}(x_j)$ are the marginals of the $x_j$, $j\in\partial i$, on the cavity graph $G^{(i)}$. Given that the cavity marginals must themselves be Gaussian, and denoting by $\omega_j^{(i)}$ their precisions, performing the $x_j$ integrals in Eq.\,\eqref{MargCav} entails that the marginal precisions $\omega_i$ are given by \cite{Ku+07, Ku08, Rog+08}
\be
\omega_i = \mu_i + \sum_{j\in\partial i} \frac{K_{ij} \omega_j^{(i)}}{K_{ij}+ \omega_j^{(i)}}\ ,
\label{Prec}
\ee
with the $\omega_j^{(i)}$ still to be determined. Following an analogous line of reasoning for the $P_j^{(i)}(x_j)$ , one can conclude that the $\omega_j^{(i)}$ must satisfy the self-consistency equations
\be
\omega_j^{(i)} = \mu_j + \sum_{\ell\in\partial j\setminus i} \frac{K_{j\ell} \omega_\ell^{(j)}}{K_{j\ell}+ \omega_\ell^{(j)}}\ .   
\label{CavPrec}
\ee
Equations \eqref{Prec} and \eqref{CavPrec} are exact on trees and become asymptotically exact on locally tree like graphs in the thermodynamic limit.

\begin{widetext}
Generalizing the line of reasoning underlying Eq.\,\eqref{MargCav} to bi-variate Gaussian marginals for two adjacent nodes $i$ and $j$ on the graph, one obtains
\bea
P_{ij}(x_i,x_j) &\propto& \exp\Big(-\frac{1}{2} (\mu_i x_i^2+ \mu_j x_j^2)-\frac{1}{2} K_{ij}(x_i-x_j)^2\Big)\nn\\
& & \times \prod_{\ell\in\partial i\setminus j} \int \rd x_\ell\,\exp\Big(-\frac{1}{2} K_{i\ell}(x_i-x_\ell)^2\Big) P_\ell^{(j)}(x_\ell)\nn\\
& & \times \prod_{\ell'\in\partial j\setminus i} \int \rd x_{\ell'}\,\exp\Big(-\frac{1}{2} K_{j\ell'}(x_j-x_{\ell'})^2\Big) P_{\ell'}^{(i)}(x_{\ell'})\ .
\label{Pij-id}
\eea
Using once more the Gaussian nature of the cavity marginals involved and performing the Gaussian integrals in Eq.\,\eqref{Pij-id} allows one, using Eq.\,\eqref{CavPrec} both for $\omega_i^{(j)}$ and $\omega_j^{(i)}$, to obtain
\be
P_{ij}(x_i,x_j) = \frac{1}{Z_{ij}} \exp\Big(-\frac{1}{2} K_{ij}(x_i-x_j)^2 - \frac{1}{2} \omega_i^{(j)} x_i^2 - \frac{1}{2} \omega_j^{(i)} x_j^2\Big)
\label{Pij}
\ee
for the joint marginal, in which 
\be
Z_{ij} = \sqrt{\frac{(2\pi)^2}{(K_{ij}+\omega_i^{(j)})(K_{ij}+\omega_j^{(i)}) - K_{ij}^2}}
\ee
is determined by normalization. After dividing this result by the marginal density $P_i(x_i)$ one obtains conditional distributions from the joint pdf Eq.\,\eqref{Pij} as
\be
P_{j}(x_j|x_i)  = \frac{1}{\sqrt{\frac{2\pi}{K_{ij} +\omega_j^{(i)}}}}\,
\exp\bigg(-\frac{1}{2} \big(K_{ij}+\omega_j^{(i)}\big) \Big(x_j -\frac{K_{ij} x_i}{K_{ij} +\omega_j^{(i)}}\Big)^2\bigg)\ .  
\ee

The above results allow one to evaluate
\be
\rho_i^h = \E_{x_i}[\chi_{\{x_i\ge h\}}] =  H(\sqrt{\omega_i} h)
\ee
and similarly
\bea
H_j(h|x_i) &=& \E_{x_j}\Big[\chi_{\{x_j\ge h\}}\Big| x_i\Big]\nn\\
 &=&  H\bigg(\sqrt{K_{ij} + \omega_j^{(i)}}\,\Big(h - \frac{K_{ij} x_i}{K_{ij} + \omega_j^{(i)}} \Big) \bigg)\ ,
\label{Hjie}
\eea
where
\be
H(z) = \int_z^\infty \frac{\rd x}{\sqrt{2\pi}} \, \exp\Big( -\frac{1}{2} x^2\Big) = \frac{1}{2} \mbox{erfc}\big(z/\sqrt 2\big)\ .
\label{Hofz}
\ee
With these results, all ingredients of the self-consistency equations \eqref{gji} as well as expressions for the local percolation probabilities $g_i$ given by Eq.\,\eqref{gi} of this paper are well defined once a solution to Eqs.\,\eqref{CavPrec} for the $\omega_j^{(i)}$ has been obtained.

\section{Asymptotic Analysis in the Vicinity of the Percolation Transition}
\label{secAsympt}
Following \cite{KuRog17}, we assume an expansion of the $g_j^{(i)}$ in Eqs.\,\eqref{gji} in powers of $0\le \varepsilon = h_c-h \ll 1$ of the form
\be
g_j^{(i)} = \varepsilon a_j^{(i)} + \varepsilon^2 b_j^{(i)} + \varepsilon^3 c_j^{(i)} + \dots
\ee
and we expand the $H_\ell(h|x_j)$ appearing on the r.h.s. of these equations in a Taylor series at $h_c$,
\be
H_\ell(h|x_j) = H_\ell(h_c|x_j) + H'_\ell(h_c|x_j) (h-h_c) + \frac{1}{2}H''_\ell(h_c|x_j) (h-h_c)^2+ \dots \ .
\ee
Inserting these expansions into Eqs.\,\eqref{gji} and collecting powers of $\varepsilon$ we obtain
\bea
\cO(\varepsilon): a_j^{(i)}\!\! &=&\!\! \sum_{\ell\in \partial_j\setminus i} B_{(ij),(j\ell)} a_\ell^{(j)}\label{O1}\\
\cO(\varepsilon^2): b_j^{(i)}\!\! &=&\!\! \sum_{\ell\in \partial_j\setminus i} B_{(ij),(j\ell)} b_\ell^{(j)} - \sum_{\ell\in \partial_j\setminus i} B'_{(ij),(j\ell)} a_\ell^{(j)}\nn\\
    & & - \frac{1}{2}\bigg[\sum_{\ell,\ell'\in \partial_j\setminus i} B^{(2)}_{(ij),(j\ell)(j\ell')} a_\ell^{(j)} a_{\ell'}^{(j)} -\sum_{\ell\in \partial_j\setminus i} B^{(2)}_{(ij),(j\ell)(j\ell)} \big(a_\ell^{(j)}\big)^2\bigg]\label{O2}\\
    \cO(\varepsilon^3): c_j^{(i)}\!\! &=&\!\! \sum_{\ell\in \partial_j\setminus i} B_{(ij),(j\ell)} c_\ell^{(j)} - \sum_{\ell\in \partial_j\setminus i} B'_{(ij),(j\ell)} b_\ell^{(j)} + \frac{1}{2}\sum_{\ell\in \partial_j\setminus i} B''_{(ij),(j\ell)} a_\ell^{(j)}\nn\\
    & & - \bigg[\sum_{\ell,\ell'\in \partial_j\setminus i}\!\!\!\! B^{(2)}_{(ij),(j\ell)(j\ell')} a_\ell^{(j)} b_{\ell'}^{(j)}
    -\sum_{\ell \in \partial_j\setminus i}\!\!\! B^{(2)}_{(ij),(j\ell)(j\ell)} a_\ell^{(j)} b_{\ell}^{(j)}\bigg] 
    + \frac{1}{3!}\bigg[\sum_{\ell,\ell',\ell''\in \partial_j\setminus i}\!\!\!\! B^{(3)}_{(ij),(j\ell)(j\ell')} a_\ell^{(j)} a_{\ell'}^{(j)} a_{\ell''}^{(j)}\nn\\
    & & -3 \sum_{\ell,\ell'\in \partial_j\setminus i} B^{(3)}_{(ij),(j\ell)(j\ell')(j\ell')} a_\ell^{(j)}\big(a_{\ell'}^{(j)}\big)^2 + 2 \sum_{\ell\in \partial_j\setminus i} B^{(3)}_{(ij),(j\ell)(j\ell)(j\ell)} \big(a_{\ell}^{(j)}\big)^3\bigg]
\label{O3}
\eea
in which 
\be
B^{(2)}_{(ij),(j\ell)(j\ell')} = \E_{x_j}\big[H_\ell(h|x_j) H_{\ell'}(h|x_j)\Big|\{x_j\ge h\}\big]\Big|_{h_c}\ .
\ee
and
\be
B^{(3)}_{(ij),(j\ell)(j\ell')(j\ell'')} = \E_{x_j}\big[H_\ell(h|x_j) H_{\ell'}(h|x_j)H_{\ell''}(h|x_j)\Big|\{x_j\ge h\}\big]\Big|_{h_c}\ .
\ee
are non-zero elements of second and third order generalizations of the non-backtracking operator defined in Eq.\,(11) of the letter, while the $B'_{(ij),(j\ell)}$ and the $B''_{(ij),(j\ell)}$ are first and second $h$-derivatives of the non-zero elements of $B$, all evaluated at $h_c$. 
    
Following \cite{KuRog17}, we can obtain a more compact representation of the above relations by introducing a $2M=\sum_i k_i$-dimensional vector $\va =\big(a_j^{(i)}\big)$. Equation \eqref{O1} then states that $\va$ is the Frobenius (right)-eigenvector of $B$ corresponding to the eigenvalue $\lambda_{\rm max}(B) = 1$. Setting  $\va = \alpha \vv$ with $\vv$ normalized to $||\vv||_1=1$ and an amplitude $\alpha$ to be determined and denoting by $\vu^T$ the corresponding Frobenius (left) eigenvector of $B$, i.e. $\vu^T = \vu^T B$, with $\vu^T \vv =1$, Eq.\,\eqref{O2} yields
\be
\alpha = - 2\frac{\vu^T B'\vv}{\vu^T B^{(2)} \vv\otimes\vv - \vu^T \tilde B^{(2)} (\vv \vv)}
\label{alpha}
\ee
Here we use the notation $(\va \vb)$ to denote a vector with components given by the product of components of $\va$ and $\vb$, i.e., $(\va \vb)_\ell^{(j)} = a_\ell^{(j)} b_\ell^{(j)}$ and an analogous construction for a vector $(\va \vb \vc)$ constructed from three vectors, and $\tilde B^{(2)}$ to denote the matrix obtained by restricting $B^{(2)}$ to be diagonal in the second pair of indices.
    
In order to obtain the $\cO(\varepsilon^2)$ contribution to the $g_j^{(i)}$ we rewrite Eq.\,\eqref{O2} as
\be
(\bbbone - B)\vb = - \alpha B'\vv - \frac{\alpha^2}{2}\Big(B^{(2)}\vv\otimes\vv -\tilde B^{(2)}(\vv\vv)\Big)\ .
\ee
It is solved by
\be 
\vb = (\bbbone - B)^+ \bigg[- \alpha B'\vv - \frac{\alpha^2}{2}\Big(B^{(2)}\vv\otimes\vv -\tilde B^{(2)}(\vv\vv)\Big)\bigg] + \beta \vv\ ,
\label{defw}
\ee
in which $(\bbbone - B)^+$ is the Moore-Penrose pseudoinverse of $(\bbbone - B)$ and $\beta$ a coefficient to be determined from Eq.\,\eqref{O3}. Indeed, writing Eq.\,\eqref{defw} as $\vb = \vw + \beta \vv$, thereby defining $\vw$, multiplying Eq.\,\eqref{O3} by $\vu^T$ yields
\bea
0 &=& -\vu^T B'\vw +\frac{\alpha}{2}\vu^T B''\vv 
    -\alpha\Big[\vu^T B^{(2)}\vv\otimes \vw - \vu^T\tilde B^{(2)}(\vv\vw)\Big]\nn\\
& & +\frac{\alpha^3}{3!}\Big[\vu^T B^{(3)}\vv\otimes\vv\otimes\vv - 3\vu^T \tilde B^{(3)}\vv\otimes(\vv\vv)
    + 2\vu^T \tilde{\tilde B}^{(3)}(\vv\vv\vv)\Big]\nn\\
& & +\beta\Big[-\vu^T B'\vv -\alpha \vu^T B^{(2)}\vv\otimes \vv + \alpha \vu^T \tilde B^{(2)}(\vv\vv)\Big]\ ,
\eea
which is easily solved for $\beta$, thereby completely determining $\vb$. Noting that Eq.\,\eqref{alpha}) can be used to simplify the coefficient of $\beta$ in the last equation to $\vu^T B'\vv$, we get 
\bea
\beta &=& \frac{1}{\vu^T B'\vv} \Bigg[\vu^T B'\vw -\frac{\alpha}{2}\vu^T B''\vv 
+ \alpha\Big[\vu^T B^{(2)}\vv\otimes \vw - \vu^T\tilde B^{(2)}(\vv\vw)\Big]\nn\\ 
& & ~~~~~ -\frac{\alpha^3}{3!}\Big[\vu^T B^{(3)}\vv\otimes\vv\otimes\vv - 3\vu^T \tilde B^{(3)}\vv\otimes(\vv\vv)
+ 2\vu^T \tilde{\tilde B}^{(3)}(\vv\vv\vv)\Big]\Bigg]
\eea

\section{Large-$c$ Asymptotics of the Percolation Threshold on $c$RRGs}
\label{secAsymptRRG}
Here we provide an asymptotic analysis of Eq.\,\eqref{hc} for the critical threshold $h_c$ of the level-set percolation transition on $c$RRGs, i.e., of
\be
(c-1) \E_{x}\Big[H(h|x)\Big|\{x\ge h\} \Big] = 1\ .
\label{hcApp}
\ee
This equation requires that 
\be
H(h|x) = H\bigg(\sqrt{K+\tilde\omega}\,\Big[h -\frac{K x}{K+\tilde\omega}\Big]\bigg)
\ee
with $\tilde\omega = \tilde\omega_+$ given by Eq.\,\eqref{tomcRRG} must be small in that part of the domain $\{x\ge h\}$ from which the $x$-expectation in Eq.\,\eqref{hcApp} derives its dominant contributions, which in turn requires $\sqrt{K+\tilde\omega}\,h \gg 1$. Assuming that in that domain we also have $K x/\sqrt{K+\tilde\omega} \ll 1$, we can perform a Taylor expansion of $H(h|x)$. To first order in $x$, it reads
\be
H(h|x) \simeq  H\Big(\sqrt{K+\tilde\omega}\,h\Big)  -\frac{Kx\,\re^{-\frac{1}{2}(K+\tilde\omega)h^2}}{\sqrt{2\pi(K+\tilde\omega)}}\ .
\label{Tay1}
\ee
This allows us to evaluate $\E_{x}[H(h|x)|\{x\ge h\}]$, using $x\sim\cN(0,1/\omega)$ with $\omega$ given by Eq.\,\eqref{omcRRG}, and thereby to rewrite Eq.\,\eqref{hcApp} to this order in the Taylor expansion as
\be
(c-1)\bigg[H\Big(\sqrt{K+\tilde\omega}\,h\Big) +\frac{1}{H(\sqrt{\omega}\,h)}\,\frac{K\,\re^{-\frac{1}{2}\omega h^2}}{\sqrt{2\pi \omega}}\,\frac{\re^{-\frac{1}{2}(K+\tilde\omega) h^2}}{\sqrt{2\pi (K+\tilde\omega))}}\bigg] \simeq 1\ .
\label{hcAT}
\ee
As our assumption $\sqrt{K+\tilde\omega}\,h \gg 1$ implies $\sqrt{\omega}\,h \gg 1$, we can use large-argument expansions for the complementary error functions in terms of which the $H(\cdot)$ appearing in Eq.\,\eqref{hcAT} are defined to rewrite Eq.\,\eqref{hcAT} as
\be
(c-1) \frac{\re^{-\frac{1}{2}(K+\tilde\omega) h^2}}{\sqrt{2\pi (K+\tilde\omega))} h} \big(1 + K h^2\big) \simeq 1
\ee
Using the shorthand $y = \frac{1}{2}(K+\tilde\omega) h^2$, we recast this into an equation for $y$, viz.,
\be
y \simeq \ln\bigg(\frac{c-1}{2\sqrt{\pi}}\bigg) -\frac{1}{2}\ln(y)+ \ln\Big(1 + \frac{2Ky}{K+\tilde\omega}\Big)\ . 
\label{yeq}
\ee
Introducing
\be
y_0 = \ln\bigg(\frac{c-1}{2\sqrt{\pi}}\bigg)
\label{defy0}
\ee
and noting that $y_0 \to \infty$ as $c\to\infty$, it is straightforward to convince oneself that the solution of Eq.\,\eqref{yeq} satisfies
\be
y = y_0 + o(y_0)\ ,
\ee
provided that $K/(K+\tilde\omega) \ll 1$ for $c\gg 1$. The solution including the $o(y_0)$ correction is most conveniently found by iteration using
\be
y_n \simeq  y_0 - \frac{1}{2}\ln(y_{n-1})  + \ln\Big(1 + \frac{2 K y_{n-1}}{K+\tilde\omega}\Big)\ ,\qquad n = 1, 2, \dots\ . 
\label{yna}
\ee
The corresponding approximations for critical values are then
\be
h_{cn} = \sqrt{\frac{2}{K+\tilde\omega} y_n}\quad \mbox{for}\quad n= 0, 1, 2, \dots\ .
\label{hna}
\ee
\end{widetext}

It remains to check, whether the assumptions made on the way are all self-consistenly satisfied. Details, including in particular the asymptotic behavior of $h_c$ itself, will clearly depend on whether the uniform edge-weight $K$ is or is not scaled with the degree $c$ of the $c$RRG. We will consider the two different conventions which are commonly made, viz., $K = K(c) \equiv 1$, and $K = K(c) = 1/c$ in turn.

For $K = K(c) \equiv 1$, we have $\tilde\omega \simeq \mu + c-2 \gg 1$ as $c \gg 1$. To satisfy $\sqrt{K+\tilde\omega}\,h \gg 1$ for $h\simeq h_c$, we thus need that $h_c \gg 1/\sqrt{K+\tilde\omega}$, which is clearly satisfied by the $h_{cn}$ of Eq.\,\eqref{hna}, given that indeed $K/(K+\tilde\omega) \ll 1$ for $c\gg 1$ in this case and therefore $y_n = y_0 + o(y_0 ) \gg 1$ for $c \gg 1$. Noting further that $\omega \simeq \mu + c-1 \gg 1$ as $c \gg 1$, and that sizeable contributions to the conditional expectation in Eq.\,\eqref{hcApp} can only come from $x = O(1/\sqrt{\omega}) \ll 1$, we also satisfy the assumption $K x/\sqrt{K+\tilde\omega} \ll 1$ on which the utility of the first order Taylor expansion \eqref{Tay1} depends. Hence, all assumptions made on the way are self-consistently satisfied and 
\be
h_{cn} \to 0\ , \qquad \mbox{as}\quad c\to \infty
\ee
in this case.

For $K = K(c) = 1/c$, on the other hand, we have $\tilde\omega \simeq \mu + \frac{c-2}{c} =O(1)$  and thus $K+\tilde\omega \simeq \mu + \frac{c-1}{c} =O(1)$ as $c \gg 1$. In order to satisfy $\sqrt{K+\tilde\omega}\,h \gg 1$ for $h\simeq h_c$, we now need that $h_c \gg 1$ for $c \gg 1$. The requirement that $K x/\sqrt{K+\tilde\omega} \ll 1$ needed for the utility of the first order Taylor expansion is now trivially satisfied for any $x = O(1)$ and thus also for $x = O(1/\sqrt{\omega})$ given that $\omega \simeq \mu + \frac{c-1}{c} =O(1)$ as $c \gg 1$. As before $K/(K+\tilde\omega) \ll 1$ for $c\gg 1$ in this case and therefore $y_n = y_0 + o(y_0 ) \to \infty$ as $c \to \infty$, hence 
\be
h_{cn} \to \infty\ , \qquad \mbox{as}\quad c\to \infty
\ee
by Eq.\,\eqref{hna}. Once more, therefore, all assumptions made on the way for the asymptotic analysis are self-consitently satisfied in this case.
\bibliography{FullGLSPerc.bbl}
\end{document}